\documentstyle[epsf]{elsart}

\begin{document}

\title{Incomplete approach to homoclinicity in a model with bent-slow manifold
geometry}

{\author S. Rajesh$^1$ and G. Ananthakrishna$^{1,2,}${\footnote{Electronic mail: garani@mrc.iisc.ernet.in}}}

\address{  $^1$ Materials Research Center,\\
 Indian Institute of Science,\\
 Bangalore - 560012.\\
 India.\\
 $^2$Center for Condensed Matter Theory,\\
 Indian Institute of Science,\\
 Bangalore - 560012.\\
 India.}

{\bf Abstract:}  
The dynamics of a model, originally proposed for a type of instability in
plastic flow, has been investigated in detail. The bifurcation portrait of
the system in two physically relevant parameters exhibits a rich variety of
dynamical behaviour, including period bubbling and period adding or Farey
sequences. The complex bifurcation sequences, characterized by Mixed Mode
Oscillations, exhibit partial features of Shilnikov and Gavrilov-Shilnikov
scenario.  Utilizing the fact that the model has disparate time scales of
dynamics, we explain the origin of the relaxation oscillations using the
geometrical structure of the bent-slow manifold. Based on a local analysis,
we calculate the maximum number of small  amplitude oscillations, $s$, 
in the periodic orbit 
of  $L^s$ type, for a given value of the control parameter. This further leads
to a scaling relation for the small amplitude oscillations. The incomplete
approach to homoclinicity is shown to be a result of the finite rate of
`softening' of the eigen values of the saddle focus fixed point. The latter
is a consequence of the physically relevant constraint of the system which
translates into the occurrence of back-to-back Hopf bifurcation.

\vspace{0.5cm}
\noindent
{\bf PACS number(s):} 05.45.Ac, 83.50.-v, 62.20.-x\\

\vspace{0.5cm}

\newpage
\section{ Introduction}
\baselineskip=0.8cm

\noindent
Number of autonomous dynamical systems exhibit complex bifurcation sequences
with alternate periodic-chaotic states in control parameter space. The
chaotic states are usually predominant mixture of the periodic states
occurring on either side of the chaotic states. These periodic states are
characterized by combinations of relatively large amplitude excursions and
small amplitude near harmonic oscillations of the trajectories and have been
referred to as mixed mode oscillations (MMOs) in the literature. The MMOs
and the accompanying complex bifurcation sequences have been observed in
models and experiments in various fields of chemical kinetics \cite
{rou84,pet92,gyo92,arn93,wan93,kop95,eps96,mil98}, electrochemical reactions 
\cite{alb89,kop92a,kop92b,arg93}, biological systems \cite{cha95}, and in
many other physical systems\cite{tom89,lef93}.\\

\noindent
Both numerical as well as analytical studies have been carried out
extensively to the explain the origin of the MMOs and the complex
bifurcation sequences these systems exhibit\cite
{alb89,kop92a,kop92b,arg93,tur81,guc90,gle84,pik81,sim82,cof86,gas84,gas87}.
Even though the origin of complex bifurcation sequences and the accompanying
MMOs may depend on the particular system under study, almost all proposed
mechanisms suggest that these bifurcation portraits are the artefact of
global bifurcations of the system \cite{guc90}. Investigations into the
global nature of the bifurcations of the MMOs and the complex bifurcation
sequences have shown that homoclinic bifurcations may be relevant to a wide
variety of systems which display the MMOs. Shilnikov\cite{shilni} has shown
that if a dynamical system possesses a homoclinic orbit which is
bi-asymptotic to a saddle focus type of equilibrium set satisfying the
Shilnikov condition, then there are countably infinite number of periodic
solutions in the vicinity of this homoclinic orbit. The analysis also shows
that in the vicinity of this homoclinic orbit, complex bifurcation sequence
can be expected in the phase portrait \cite{guc90,gle84}.\\

\noindent
Another approach to explain the complex bifurcation sequences has been
through Gavrilov - Shilnikov scenario. It has been shown that systems having
homoclinic tangencies to periodic solutions possess quasi-random dynamics
and MMO like behavior in the control parameter space\cite{gav72}. Each of
these scenarios are characterized by bifurcation diagrams obtained from the
stability analysis of the homoclinic orbit and by the corresponding scaling
relations involved in the approach to the homoclinicity. Apart from these
studies on the homoclinic bifurcations of the continuous time systems,
attempts have also been made to study the MMOs through discrete maps \cite
{pik81,sim82,cof86}.\\

In order to understand the numerically obtained Poincare maps from such
model systems and those obtained from experiments, attempts have been made
to derive map structure starting from local analysis\cite{gle84,gas84,gas87}. Such attempts have been reasonably successful in the sense that the
features predicted by the derived maps agree with the features of the
numerically obtained Poincare maps. However, due to the very nature of
global bifurcations, there is no easy way of classifying the entire complex
bifurcation sequences of these systems. Hence numerical evidence plays a
crucial role in classifying the dynamics of these systems.\\

\noindent
Yet another approach to the understanding of these complex bifurcation
sequences and the MMOs is based on the analysis of the structure of the slow
manifold \cite{arn93,kop92a,kop92b,bar88,hau96,den94} of the system of
equations. A standard slow manifold structure that has been used in the
study of MMOs is the $S-$shaped structure\cite{rossler,boi76} wherein the
upper and lower pleats are attractive and the middle branch is repulsive. By
appropriately locating the fixed point on the upper or lower pleat, the
origin of MMOs has been explained\cite{bar88}. In this construction of slow
manifold, Shilnikov's criterion is satisfied if the direction of approach of
the fast variable is transverse to the slow manifold containing the fixed
point. However, in many situations\cite{kop95,kop92a,kop92b}, including the
system under study, the approach to the saddle fixed point by the fast
variable is not transversal. Depending on the nature of the approach of the
fast variable to the fixed point, a classification for the MMOs has been
suggested by Koper {\it et. al.} \cite{kop92b} as type-I and type-II
corresponding to the tangential approach and transversal approach (to the
slow manifold) respectively. It has also been suggested that the occurrence
of MMOs and incomplete approach to homoclinicity is related to the presence
of Hopf bifurcation close to the fold of the S-shaped slow manifold structure \cite{kop92a}.\\

\noindent
In this paper, we analyze the complex dynamical behavior of a model which
has been introduced in the context of a type of plastic instability called
the Portevin - Le Chatelier effect. The model exhibits a rich variety of
dynamics such as period bubbling, doubling and the complex bifurcation
sequences. Two distinct features of the model are the atypical nature of the
relaxation oscillation and the MMOs. The latter exhibits partial features of
Shilnikov, and Gavrilov-Shilnikov scenario and shows a incomplete approach
to the homoclinic point. 
Our effort is focussed in understanding this issue in the context of our
model. To begin with we study the nature and the origin of the relaxation
oscillations by analyzing the geometry of the slow manifold which controls
the relaxation oscillations. We show that the underlying cause of the
relaxation oscillations is due to the atypical bent geometry of the slow
manifold. This feature forms the basis of further analysis of the nature of
the MMO sequences, and the incomplete approach to homoclinic bifurcation.
The paper is organized as follows. In section II, for the sake of
completeness, we start with a brief introduction to the phenomenon followed
by a description of the model. Section III contains a detailed analysis of
the complex bifurcation sequences exhibited by the model in the plane of two
physically interesting parameters. Section IV contains a discussion on the
origin of relaxation oscillations using the geometry of the slow manifold.
Using this, we explain the origin of the MMOs and derive a scaling relation
involving the maximum number of periodic orbits allowed for given value of
the control parameter. The analysis helps us to understand the cause of the
incomplete approach to homoclinicity. We conclude the paper with discussion
and conclusions in section V.

\section{ A Dynamical Model for Jerky Flow}

\noindent
Since the model is rooted in the area of plastic instability, for the sake
of completeness, we start with a brief introduction to the phenomenon. The
Portevin-Le Chatelier (PLC) effect is a plastic instability manifesting when specimens of metallic
alloys are deformed under tensile deformation. Under normal conditions, the
stress-strain curve is smooth. However, repeated yield drops occur when the
material parameters are in the regime of instability. Each of the load drops
is related to the formation and propagation of dislocation bands \cite
{bri70,kub93}. The PLC effect (or the jerky flow) is seen in several
metallic alloys such as commercial aluminium, brass, alloys of aluminium and
magnesium \cite{bri70}. The phenomenon is observed only in a window of
strain rates and temperature. It is generally agreed that the microscopic
cause of the instability is due to the interaction of dislocations with
mobile point defects. This leads to the negative strain rate characteristic
of the yield stress. The basic idea was formulated by Cottrell \cite{cot53}
few decades ago. However, this model and its extensions do not deal with the
time dependent nature intrinsic to the phenomena.\\

\noindent
The first dynamical description was attempted Ananthakrishna and coworkers
several years ago \cite{ana81b,ana82a}. The basic idea of the model 
is that most of the 
generic features of the PLC effect stem from nonlinear interactions between 
defect populations. The model in its  original form does introduce 
spatial dependence of specific nature. However, further analysis of 
the model ignores 
the details of spatial inhomogeneous structure. The model consists of three 
types of dislocations and some transformations between them. The model has 
proved to be very successful in that it could explain most of the experimentally 
observed features such as the existence of bounds on the strain rate for the 
PLC effect to occur, the negative strain rate sensitivity, etc.,\cite{ana82a,val83,ananl}. Several
aspects of the model has been investigated \cite{mul97,gla97,mar98}.\\
   
A few comments may be in order regarding the spatial aspect of the PLC effect.
The nature of spatial terms that should be introduced in the 
description of the PLC effect  has been a controversial  topic\cite{Kubin99}. 
However, there is some consensus that 
the double cross slip mechanism plays an important role in the spatial 
aspect. In the above model, justification for ignoring  the spatial
inhomogeneous structure  and considering only the temporal  aspects of the phenomenon 
is that the variables (dislocation densities) correspond to the collective 
degrees of freedom of the spatially extended systems. But, if one were to be 
interested in spatial aspects directly, we refer the reader to an improved 
version of the model\cite{ana93}.  Further work to improve the model by 
including the nonlocal effects of immobile density along with the cross 
slip term is under active investigation.\\

\noindent
One important prediction of the model is that the phenomena should be
chaotic in a certain regime of applied strain rate\cite{ana83,ananl,ana90}.
This prediction has been verified by analyzing stress signals obtained from
single and polycrystalline samples \cite{ana93,ana95,ananl3,nor96,nor97}.
Further, the number of degrees of freedom required for a dynamical description 
of the phenomenon estimated from the analysis was
found to be four or five consistent with that envisaged in the model. 
Moreover, since the physical system is spatially extended, this 
reduction to few degrees of freedom does suggest that these modes correspond 
to collective degrees of freedom of the participating defects. From 
this point of view, dealing with the temporal aspect appears to be justified. Therefore,
it is natural to investigate the chaotic behavior of the model in its own
right. Some preliminary results on the chaotic aspects of the model have
been published earlier \cite{ana83,ananl,ana90}.\\

\noindent
The model consists of mobile dislocations and immobile dislocations and
another type which mimics the Cottrell's type, which are dislocations with
clouds of solute atoms \cite{ana82a}. Let the corresponding densities be $N_m
$, $N_{im}$ and $N_i$, respectively. The rate equations for the densities of
dislocations are: 
\begin{eqnarray} 
\dot{N}_m & = & \theta V_m N_m - \beta N_m^2 -\beta N_m N_{im} +\gamma N_{im}\nonumber\\ 
                           & & \hspace{4.2cm}-\alpha_m N_m\, \\
\dot{N}_{im} & = &\beta N^2_m -\beta N_{im} N_m -\gamma N_{im} +\alpha_i N_i, \\
\dot{N}_i & = & \alpha_m N_m - \alpha_i N_i.  
\end{eqnarray}
The overdot, here, refers to the time derivative. The first term in Eq. (1)
is the rate of production of dislocations due to cross glide with a rate
constant $\theta$. $V_m$ is the velocity of the mobile dislocations which in
general depends on some power of the applied stress $\sigma_a$. The second
and third term refer to annihilation or immobilization processes. The fourth
term represents the remobilization of the immobile dislocations due to
stress or thermal activation (see the loss term $\gamma N_{im}$ in Eq. 2).
The last term represents the immobilization of mobile dislocations either
due to solute atoms or due to other pinning centers. $\alpha_m$ refers to
the concentration of the solute atoms which participate in slowing down the
mobile dislocations. Once a mobile dislocation starts acquiring solute atoms
we regard it as a new type of dislocation, namely the Cottrell's type $N_i$.
This process is represented as a gain term in Eq. (3). As they acquire more
and more solute atoms they will slow down and eventually stop the
dislocation entirely. At this point, they are considered to have transformed
to $N_{im}$. This process has been represented by the loss term in Eq. (3)
and a gain term in Eq. (2).

These equations should be dynamically coupled to the machine equations
describing the rate of change of the stress developed in the sample. This is
given by 
\begin{eqnarray}
                        \dot{\sigma_a} = \kappa(\dot{\epsilon_a} - B_0N_mV_m),  
\end{eqnarray}
\noindent
where $\kappa$ is the effective modulus of the system, $\dot{\epsilon}_a$ is
the applied strain rate, 
$V_m$ is the velocity of mobile dislocations and $B_0$ is the Burgers
vector. These equations can be cast into a dimensionless form by using the
scaled variables 
\begin{eqnarray*}
x = N_m \left(\frac{\beta}{\gamma}\right),
y = N_{im}\left(\frac{\beta}{\theta V_0}\right),\\
z = N_i \left(\frac{\beta\alpha_i}{\gamma\alpha_m}\right), \tau=\theta V_0 t,
\hbox{\,\,and\,\,} \phi=\left({\sigma_a\over{\sigma_0}}\right). 
\end{eqnarray*}
Using the power law dependence $V_m=V_0({\frac{\sigma_a}{\sigma_0}})^m $,
Eqs. (1-3) and (4) can be rewritten as 
\begin{eqnarray}
\dot{x} & = & \phi^mx - ax -b_0x^2 -xy +y,
\\
\dot{y} & = & b_0\left(b_0x^2 -xy-y+az\right),
\\
\dot{z} & = & c(x-z),
\\
\dot{\phi} & = & d\left(e-\phi^mx\right),
\end{eqnarray}
\noindent
Here $a = (\alpha_m/\theta V_0), b_0=(\gamma/\theta V_0), c=(\alpha_i/\theta
V_0)$, $\kappa=(\theta \beta \sigma_0d/\gamma B_0)$ and 
$e=(\dot{\epsilon}_a\beta/B_0 V_0\gamma)$. For these set of equations 
there is only one steady
state which is stable. There is a range of the parameters $a,b_0,c,d,m$ and $e$
for which the linearized equations are unstable. In this range $x,y,z$ and $%
\phi$ are oscillatory.\\

\noindent
Among these physically relevant parameters, we report here the behavior of
the model as a function of most important parameters namely the applied
strain rate $e$ and the velocity exponent $m$. We use $e$ as the primary
control parameter for the analysis. The values of other parameters are kept
fixed at $a=0.7,b_0=0.002,c=0.008,d=0.0001$ and $k=1.0$. The present choice of
parameters {\it does not necessarily correspond to a realistic experimental
situation,} although there is a range of allowed values. As can be verified
these equations exhibit a strong volume contraction in the four dimensional
phase space. We note that there are widely differing time scales
corresponding to $a,b_0,c$ and $d$ (in the decreasing order) in the dynamics
of the model. For this reason, the equations are stiff and numerical
integration routines were designed specifically to solve this set of
equations. We have used a variable order Taylor series expansion method as
the basic integration technique with coefficients being determined using a
recursive algorithm. Most of the bifurcation analyses were performed using
these indigenous routines. AUTO software package\cite{AUTO} was used
exclusively for two parameter continuation of bifurcation points.
\newpage
\section{Summary of bifurcation exhibited the model}

\noindent
In an attempt to understand the complex bifurcation sequences exhibited by
the model, we start with an outline of the gross features of the phase
diagram in the $(m,e)$ plane shown in Fig. 1. In our discussion, we consider 
$m$ to be the unfolding parameter. For values of $m > m_d \sim 6.8$, the
equilibrium fixed point of the system of equations is stable. At $m=m_d$, we
have a degenerate Hopf bifurcation as a function of $e$. For values less
than $m_d$, we have a back-to-back Hopf bifurcation. The periodic orbit
connecting these two Hopf bifurcations is referred to as the Principal
Periodic Orbit (PPO). The dynamics of the system is essentially bounded by
these two Hopf bifurcations. In Fig. 1, the broken line represents the Hopf
bifurcation and the dotted lines correspond to the first three successive
period doubling bifurcations leading to period 2, 4 and 8 orbits. The region
in between the first period two and the Hopf bifurcation line exhibits
monoperiodic relaxation like oscillations.\\

The PPO for most of the parameter plane $(m,e)$ is born through a
subcritical Hopf bifurcation leading to relaxation like oscillations.
However, the narrow region between the Hopf bifurcation line (corresponding
to large values of $e$) and the period two regime is characterized by small
amplitude nearly symmetric coplanar limit cycles. The complex bifurcation
sequences, characterized by alternate periodic-chaotic sequences seen in the
parameter space, are roughly indicated by the hatched region. Since the part
of the hatched region extends beyond the outermost period doubling line
(large values of $e$ and small values of $m$), both the MMOs and the small
amplitude monoperiodic limit cycle solutions coexist in this region. (See
also Fig. 3.) A codimension two bifurcation point in the form of a cusp
(shown as a filled diamond) at $(e_c,m_c)$ formed by merging of the locus of
two saddle node periodic orbits (represented by bold lines) of the PPO is
also displayed in Fig. 1. Apart from these bifurcations, we failed to detect
any other bifurcation or equilibrium set in the phase space. We will deal
with each of these regions in detail below.\\

\noindent
Bifurcation diagrams have been obtained by plotting the maxima of any one of
the variables $x$,$y$,$z$ or $\phi$ as a function of the control 
parameters ($e,m$). We have mostly shown the bifurcation diagrams in the variable $x$.
This choice enables a good visual representation of the bifurcation diagram
since the maxima of the $x$ variable is quite large compared to other
variables. Based on the nature of the bifurcation sequences, the parameter
space can be broadly grouped into two regions, viz. $m\geq 2.0$ and $m<2.0$,
and we will discuss the changes as a function of $e$ fixing $m$ at a
particular value.

\subsection{Region $m \geq 2.0$}

\noindent
We briefly summarize the results for this region. For values of $m>2.16$,
the bifurcation diagrams are characterized by an incomplete period doubling
cascades followed by reverse period doubling bifurcations, displayed as
nested bubbles of periodic states (see Fig. 2). As $m$ decreases, the number
of periodic bubbles nested in the structure increases as $2^n$, with $%
n\rightarrow \infty $, culminating in chaos. Just below $m=2.16$, the
disjoint chaotic bubbles collide with each other forming an extended
attractor which has been referred to as 'bubble bursting' in the literature%
\cite{kno83}. Similar features have been observed with $m$ as the control
parameter keeping $e$ fixed at an appropriate value. The rates of the period
doubling (PD) bifurcations as well as the reverse period doubling
bifurcations with respect to $e$ and $m$ fall close to the value of the
Feigenbaum's constant for quadratic unimodal maps, namely $\delta _F=4.66$.
When the value of $m$ reaches a critical value $m=2.11$, a period three
cycle is born through a saddle node bifurcation and has the largest width
(in $e$) for $m\sim 2.0$ in this regime.

\subsection{Region $m < 2.0$}

\noindent
For $m<2.0$, the system exhibits qualitatively different behavior compared
to $m>2.0$, in the sense that higher order MMOs emerge gradually as $m$ is
decreased. In this region, the system exhibits complex bifurcation sequences
or the alternate periodic-chaotic sequences which are characteristic to this
system. The stable periodic orbits in the bifurcation sequence typically
exhibit MMO nature and they are labelled by $L^s$, where $L$ is the number
of large amplitude loops and $s$ is the number of small amplitude loops of
the periodic orbit. These MMOs are heralded by the creation of the period
three ($L^s:1^2$) region after the PD cascade to chaos in the bifurcation
plot. To illustrate the nature of bifurcations in this region, we fix the
unfolding parameter at $m=1.8$ and discuss the bifurcation with respect to $e
$.\\

\noindent
Figure 3 shows the bifurcation sequence with alternating chaotic and
periodic states along with the higher order periodic isolas (isolated
bifurcation curves). The unstable periodic orbits are shown by dashed lines.
In the case of isolas, we have shown only the largest amplitude isola for
any given periodicity. The first instability of the PPO through a PD
bifurcation opens up a period doubled orbit having a large parameter width
in $e$. This feature persists for the entire $m_c<m<2.0$ regime. As in the
case of $m=2.0$, a period three isola is born through a saddle node (SN)
bifurcation from the chaotic attractor. In Fig. 3, stable periodic orbits
are shown to be bounded between PD bifurcation (filled circle) and the SN
bifurcation (filled triangle). The sequential way SN and PD bifurcations
arrange themselves to form an isola can be easily understood by considering
the behavior of the Floquet eigenvalue of the periodic orbit. The period
three orbit is born in a SN bifurcation accompanied by the disappearance of
the first chaotic window attractor. At this value of $e$, Floquet eigenvalue
is at $+1$ creating a pair of stable and unstable period three orbits. As $e$
is increased, the eigenvalue of the unstable periodic orbit increases beyond 
$+1$ while the eigenvalue corresponding to the stable orbit keeps decreasing
and crosses $-1$ resulting in a PD bifurcation. Due to the isola structure,
further increase in $e$ makes this eigenvalue cross back $-1$ thus
restabilizing the period three orbit. For any further increase in $e$, the
Floquet eigen values of the stable and unstable orbit merge again at $+1$
and vanish in another SN bifurcation to complete the isola structure. The
next higher order isola is also created through a SN resulting in the
disappearance of the chaotic attractor born from the destabilisation of
stable period three orbit. Higher order periodic orbits (isolas) are formed
in a similar way. Note that these isolas are independent of the PPO. We
refer to this sequence of periodic orbits of this form as the principal
period adding sequence (PPAS) or the principal Farey sequence. In the MMO
notation, the PPAS can be written as $L^s:1^n$ where $n=2,3,4,...$. As the
periodicity increases, the width of the isolas in $e$ decreases. Since the
isolas are independent of the PPO, any change in the stability of the PPO
has no effect on the nature of the sequence. This is evident from the
bifurcation diagram where higher periodic orbits (isolas) are seen even
after stability of the PPO is reestablished. The restabilisation of the PPO
is through a reverse period doubling cascade from chaos. This chaotic
segment formed from the reverse period doubling of the PPO expands in an
interior crisis due to its collision with an unstable periodic orbit of the
next higher period isola (period three in the case of $m=1.8$). This crisis
point, shown by an arrow, marks the lower boundary of the multistability
region in the bifurcation diagram.\\

\noindent
The inset of Fig. 1 shows the expanded region of interest of the phase
diagram in the $(m,e)$ plane. As can be seen, the locus of the SN
bifurcations corresponding to the MMOs are distinct and higher period SN
bifurcation curves cross the lower ones resulting in the period $n$ isola
extending beyond the period ($n-1$) isola. Moreover, the region where SN
bifurcation curves overlap with the region of the period doubling curves,
multistability regions in the parameter space $(m,e)$ are created. This is
clearly seen in Fig. 3 where bifurcation diagram for $m=1.8$ is shown. (See
also Fig. 4.) Typically, the same mechanism as described for the case of $%
m=1.8$ operates for the period adding sequences in the region $2.0>m>m_c$,
where $m_c\sim1.1$ is the value of $m$ at the cusp point. As $m$ decreases
from $m=2.0$, higher number of stable periodic windows are accommodated with
concomitant decrease in the width of the chaotic regimes separating the
periodic windows. The arithmetically increasing periods of the orbits going
from left to right form an incomplete period adding sequence with decreasing
widths for higher order periodic windows. These features are shown in the
bifurcation diagrams for $m = 1.8$, and $m = 1.2$ in Fig. 3 and Fig. 4
respectively.\\

Below $m_c$, the bifurcation diagram is even more interesting and rich.
Here, we outline the features related to the Farey states. For the case $%
m=1.0$, only three principal Farey states denoted by $L^s$, $s=1,2$ and $3$
survive, as shown in Fig. 5. The well developed sub-Farey sequences are also
shown in the inset of Fig. 5. The sub-Farey states created go from right to
left in contrast to the principal Farey states (see Fig. 5). All these
sub-Farey sequences culminate in a SN bifurcation. While in the first
bifurcation of the PPO (SN bifurcation), the transition is from $1^0
\rightarrow \infty^1$, the mid region of parameter accommodates both the
large amplitude and small amplitude solutions with nearly equal weights.
Towards the end region of $e$, we find no fine structure typical of the
first two principal chaotic windows.
\newpage
\section{Mechanism of relaxation oscillations}

\noindent 
One characteristic feature of the dynamics of the system is its strong
relaxation nature. This is seen even in the case of the mono-periodic
solutions emerging from the Hopf bifurcation for small values of $e$. This
feature, of course persists for other regions of the $(m,e)$ plane where the
MMO type of oscillations are also seen. These two aspects are interrelated
and are a result of the structure of the slow manifold as we will show.
Since our system does not follow the known homoclinic scenarios, we look for
a new mechanism for the MMOs based on the mechanism proposed for the
relaxation oscillations.

\subsection{Relaxation oscillations}

\noindent
Relaxation oscillations are highly nonlinear oscillations with large
amplitude excursions of the fast variable. These oscillations arise as a
consequence of the existence of a fast time scales compared to the time
scales of other variables in the dynamics of the system. The relaxation
oscillations have been an intense area of research in the context of
biological rhythms\cite{ber95}. The relaxation oscillations that manifest in
the model under study is a type of relaxation oscillation wherein the fast
variable takes on large values for a short time after which it assumes small
values of the same order of magnitude as that of the slow variables. The
time spent by the fast variable in the part of phase space where the
amplitude of the fast variable is small is a substantial portion of the
period of the orbit. It is this type of relaxation oscillation that is
dominantly seen in our system, even though other types of relaxation
oscillations are also seen \cite{val83} for certain other regimes of the $%
(m,e)$ plane. We shall refer to this type of relaxation oscillation as
pulsed type relaxation (PTR) oscillation. A typical plot of $x(t)$ is shown
in inset of Fig. 6 for $e=200.0$ and $m=1.2$.\\

In the case when two disparate time scales are present in the dynamics,
using multiple scale perturbative analysis, Baer and Ernaux have shown that
Hopf bifurcation can lead to relaxation type of oscillations\cite{bae86}.
They have shown that nearly sinusoidal solutions born out of the Hopf
bifurcation change over to relaxation like oscillations in a small region of
the value of the control parameter. In such a case, the crossover to
relaxation oscillations is confined to the slow manifold around the fixed
point. As we will see, the nature of the relaxation oscillation in our model
is very different from that discussed by Baer and Ernaux. Here, it suffices
to say that the PTR is a result of the evolution wherein the trajectories
visiting the slow manifold region around the fixed point are pushed out to
another part of the slow manifold away from the fixed point where
trajectories spend considerable fraction of its period.\\

\noindent
To understand the nature of the relaxation oscillations, we first study the
structure of the slow manifold ($S$) and the behavior of the trajectories
visiting different regions of $S$. Consider the slow manifold given by 
\begin{eqnarray}
                     \dot{x} = g(x,y,\phi) &=& -b_0 x^2 + x \delta + y = 0 
\end{eqnarray}
\noindent
with $\delta =\phi ^m-y-a$. Here, the slow variables $y$ and $\phi $ (and
therefore $\delta $) are regarded as parameters. Further, as we will see
below, it is simpler to deal with the structure of the slow manifold in
terms of the $\delta $ instead of both $y$ and $\phi $. Then, the physically
allowed solution of the above equation is 
\begin{equation}
x=\frac{\delta +\sqrt{\delta ^2+4b_0y}}{2b_0}
\end{equation}
\noindent
where $\delta $ can take on both positive and negative values. Noting that $%
b_0$ is small and therefore $\delta ^2\gg 4b_0y$, two distinct cases arise
corresponding to $\delta >0$ and $\delta <0$ for which $x\sim {\delta }/{b_0}
$ and $x\sim -{y}/{\delta }$ respectively. Further, since the slow variable $%
\phi $ and $y$ take on values of the order of unity, the range of $\delta
=\delta (y,\phi )$ is of the same order as that of $\phi $ and $y$ (as is
evident from Figs. 6 and 7). Thus, we see that $x\sim -{y}/{\delta }$ is
small and $x\sim {\delta }/{b_0}$ is large. For values around $\delta =0$
and positive, we get $x\sim \left( {y}/{b_0}\right) ^{1/2}$.\\

\noindent
Let $S_1$ denote the region of slow manifold values of $x$ corresponding to $%
\delta >0$ and $S_2$ the region of slowmanifold values of $x$ for $\delta <0$%
. The bent-slow manifold structure along with the two portions of the slow
manifold are shown by a bold lines in the $(x,\delta )$ plane in Fig. 6. A
local stability analysis for points on $S_1$ and $S_2$ shows that $\partial
g/\partial x=\delta -2b_0x$ is negative. This implies that the rate of
growth of $x$ is damped and hence these regions, $S_1$ and $S_2$ will be
referred to as attracting. In Fig. 6, we have shown a trajectory
corresponding to a mono-periodic relaxation oscillation ($m=1.2$ and $e=200.0
$) by a thin line. As can be seen, the trajectory spends most of the time on 
$S_1$ and $S_2$. For points below the line $2b_0x=\delta $ ($\delta >0$), $%
\partial g/\partial x>0$ implying a positive rate of growth of the $x$
variable and hence we call this region as repulsive or 'unstable' (shaded
region of Fig. 6). We stress here that this region is not a part of the slow
manifold. Even then, the trajectory starting on $S_2$ does continue in the
direction of increasing $\delta $ beyond $\delta =0$. Once the trajectory is
in this region, it moves up rapidly in the $x$ direction (due to the
`unstable' nature) until it reaches $x={\delta }/{2b_0}$ line, thereafter,
the trajectory quickly settles down on to the $S_1$ part of the slow
manifold due to the fact ${\partial g}/{\partial x}$ becomes negative. As
the trajectory descends on $S_1$ approaching $S_2$, we see that the
trajectory deviates away from $S_1$. This happens when the value of $x$ is
such that $2b_0x<\delta $, i.e., ${\partial g}/{\partial x}>0$. Thus, points
on $S_1$ satisfying this condition are locally unstable. (For points in this
neighborhood $\delta \sim 0.2,x\sim 50$.) Thus, the trajectory makes a jump
from $S_1$ to $S_2$ in a short time. This roughly explains the origin of the
relaxation oscillation in terms of the reduced variables $\delta $ and $x$.\\

\noindent
The actual dynamics is in a higher dimensional space and a proper
understanding will involve the analysis of the movement of the trajectory in
the appropriate space. Moreover, quite unlike the standard $S-$ shaped
manifold with upper and lower attracting pleats with the repulsive
(unstable) branch, in our model, both branches of the bent-slow manifold are
connected, and there is no repulsive branch of the slow manifold. {\it Thus,
the mechanism of jumping of the orbit from $S_2$ to $S_1$ is not clear.} In
order to understand this, consider a 3-d plot of the trajectory shown in
Fig. 7. Retaining the same notation for the 3-d regions of the slow manifold
as that used for the $x - \delta$ plane, regions $S_1$ and $S_2$ are shown
in Fig. 7. As can be seen, the region $S_2$ corresponding to small values of 
$x$ lies more or less on the $y-\phi$ plane and the region $S_1$
corresponding to large values of $x$ is nearly normal to the $y-\phi$ plane
due to the large $b_0^{-1}$ factor. (Note that  the scales of $y$ and $\phi$
are the smallest for the system. ) Regions $S_1$ and $S_2$ are demarcated by
the `fold curve' which lies in the $y-\phi$ plane and is given by $\delta =
\phi^m-y-a = 0$.  As in the case of $x-\delta$ plane, in 3-d space also, the
rapidly growing nature of the trajectory seen in the approximate region
below the surface of $2b_0x=\phi^m-y-a$ and lying to right of the `fold
curve' is due to $\partial g/\partial x > 0$.\\

\noindent
The principal features of the relaxation oscillations that we need to
explain are: a) very slow time scale for evolution on $S_2$, b) fast
transition from $S_2$ to $S_1$ and c) evolution on $S_1$. As mentioned in
the introduction, these are related to the slow - fast time scales. In order
to understand this, we shall analyze Eqs. (6) and (8) by recasting them in
terms of $\delta $. In the whole analysis it would be helpful to keep in
mind the range of values of $x,y,z$ and $\phi $ (shown in Figs. 6 and 7), in
particular, their values as the trajectory enters and leaves $S_1$. Consider
rewriting Eq. (6) valid on the slow manifold $S$ in terms of $\delta $: 
\begin{equation}
\dot y=b_0(x\delta -xy+az).
\end{equation}

\noindent
The idea is to study this equation along with Eq.(8) in specific regions of
the phase space to understand the general features of the flow, viz., on $S_2
$, just outside $S_2$, and on $S_1$. The presence of the $z$ variable in Eq.
(11) poses some problems. However, it is possible to get a rough estimate of
the magnitude of $z$ and the relative changes in the values of $z$ which is
all that will be needed for our further discussions. To see this, consider
Eq. (7) from which we see that $z$ follows $x$. Further, Once the trajectory
moves out of $S_2$, $x$ changes rapidly and therefore the value of $z$
increases (in a relatively short interval of time), reaching its maximum
value, $z_{max}$, just before the trajectory returns to $S_2$. When the
trajectory is on $S_2$, since $x \sim y/\vert \delta \vert$,  from Eq. (7)
we see that the value of $z$ is slowly decreasing (with a time constant $%
c^{-1}$) starting from $z_{max}$, reaching its minimum value, say $z_{min}$,
around the time when the trajectory leaves $S_2$. In other words, the
magnitude of $z$ is maximum when the trajectory enters $S_2$ and minimum
when it leaves $S_2$. Further, we note that $x = - y/\delta \ll x_0 = z_0
\sim e/2$ and $z$ oscillates around its equilibrium value $z_0$. Thus, the
values of $z_{max}$ and $z_{min}$ are larger than the range of allowed
values of $y$. (Note that this is also consistent with the fact that the
time scale of $z$ is larger than that for $y$ and $\phi$.)\\

\noindent
Consider the behavior of Eq. (11) on $S_2$. Using the values of $x\sim {y}/{%
\left| \delta \right| }$, we get 
\begin{equation}
\dot y=b_0\left[ -y-\frac{y^2}{\left| \delta \right| }+az\right] .
\end{equation}

\noindent
By noting that on $S_2$, $z$ decreases from $z_{max}$ to $z_{min}$, we see
that there is a range of small values of $y$ for which $\dot{y} > 0$ and for
relatively larger values of $y$, $\dot{y} < 0$. Thus, $y$ clearly has a
turning point on $S_2$ beyond which $y$ decreases. \\

\noindent
Next, consider the changes in $\phi $. Using the value of $x={y}/{|\delta |}$
on $S_2$ in Eq. (8), we find that $e$ is much larger than ${\phi ^my}/{%
\left| \delta \right| }$, since these variables are of the order of unity.
Thus, $\phi $ increases linearly, at a rate close to $de<<1$. Considering
the fact that $\dot x\sim 0$ for the entire interval the trajectory is on $%
S_2$, the time scale of evolution of the trajectory is entirely controlled
by the two slow time scales of $y$ and $\phi $. This roughly explains the
behavior of the trajectory on $S_2$.\\

\noindent
Now consider the behavior of Eq. (11) in a small region just outside $\delta
=0$. Using $x\sim \left( {y}/{b_0}\right) ^{1/2}$ valid for $\delta \ge 0$,
we get, 
\begin{equation}
\dot y=b_0\left[ \left( \frac y{b_0}\right) ^{1/2}(\delta -y)+az\right]
\approx b_0\left[ \frac{-y^{3/2}}{b_0^{1/2}}+az\right] .
\end{equation}

\noindent
We are interested in investigating the behavior of Eq. (13), for $z=z_{max}$%
, appropriate as the trajectory approaches $S_2$ and $z=z_{min}$,
appropriate as the trajectory leaves $S_2$. Consider the first choice $%
(z=z_{max})$ corresponding to the trajectory as it approaching $S_2$ from
outside ($\delta >0$). Then, using the value of $b_0$, an order of magnitude
calculation shows that there is a range of small values of $y$ for which $%
\dot y>0$. This implies that $y$ grows for small $y$, meaning that the
trajectory moves towards $S_2$. Now consider using $z=z_{min}$ corresponding
to the situation when trajectory has left $S_2$. Similar estimation shows
that there is a range of (relatively larger) values of $y$ for which $\dot
y<0$. This implies that $y$ decreases for relatively large values of $y$,
meaning that the trajectory is moving away from $S_2$. ( Note that for this
case, there may or may not be a range of $y$ for which $\dot y>0$.) Thus, in
both cases the directions of growth of $y$ for small and large $y$ just
outside $S_2$ are consistent with the behavior of $y$ just inside $S_2$.
(See Fig. 7.)\\

\noindent
Now, consider Eq. (8) with $x \sim \left( {y}/{b_0} \right)^{1/2}$ valid for
the region $\delta$ positive but small. Then, 
\begin{equation}
\dot{\phi} = d\left[ e - \phi^m \left( \frac{y}{b_0} \right)^{1/2}\right]. 
\end{equation}

\noindent
Keeping in mind the order of magnitude of $b_0$, and the fact that $y$ and $%
\phi$ are of the order of unity, the magnitude of $\phi^m\left( {y}/{b_0}%
\right)^{1/2}$ is seen to be larger than its value on $S_2$. Note that a
quick order of magnitude calculation shows that there are values of $y$ and $%
\phi$ such that $\phi^m \left( {y}/{b_0}\right)^{1/2}$ is of the order of $e$
which implies that $\phi$ is about to decrease and therefore is near its
maximum. Moreover, if anything, $\phi^m x$ in Eq. (8) increases as the
trajectory tends to moves out of $S_2$, since $\dot{x} \sim x\delta$ just
outside $S_2$. This implies $\phi$ will eventually decrease.\\

\noindent
Combining the results on $\dot y$ and $\dot \phi $ for regions just outside
and inside the `fold', we see that the trajectory enters $S_2$ in the region
corresponding to small values of $y$ and $\phi $, and makes an exit for
relatively larger values of $\phi $ and $y$ (compared to their values as the
trajectory enters $S_2$). Finally, we can see that just to the right of $%
\delta =0$ line, $\dot x\sim x\delta $, with $\delta $ very small, which
suggests that the time constant is small. Thus, the growth of $x$ is slow in
the neighborhood of $\delta =0$, and is tangential to the $S_2$ plane even
in the `unstable' region. However, once the trajectory moves away from $%
\delta =0$, the growth of the trajectory is controlled by $\partial
g/\partial x$ and hence the time scale of growth of $x$ is of the order of $%
\delta ^{-1}$ which is of the order of unity. This essentially explains why
the trajectory tends to move into the `unstable' region and grows rapidly.\\

\noindent
Once in the `unstable' region, the value of $x$ continues to grow in this
region of the phase space as can be seen from Eq. (8) until the value of $x$
is such that $\phi^mx=e$ is satisfied. Beyond this value of $\phi$, $\dot
\phi$ is negative. Thus, the trajectory leaving $S_2$ eventually falls onto
the $S_1$ part of the slow manifold. We can again evaluate $\dot{y}$ and $%
\dot{\phi}$ just as the trajectory reaches $S_1$. Using $x \sim {\delta}/{b_0%
}$ in Eq. (11), we find 
\begin{equation}
\dot{y} = b_0 \left[ \frac{\delta}{b_0} (\delta - y) + a z \right], 
\end{equation}
\noindent
The sign of $\dot{y}$ is determined by the factor ($\delta - y$) at the
point where the trajectory reaches $S_1$. To see the relative magnitudes of $%
\delta$ and $y$, consider obtaining an equation for $\delta$ starting from $%
g(x,y,\delta) = 0$. Differentiating this and using $\dot{x}=0$ on for the
slow manifold, we get

\begin{equation}
\dot{\delta} = -\frac{b_0}{x} \left[x(\delta - y) + a z\right]. 
\end{equation}

\noindent
Using $x \sim {\delta}/{b_0}$ on $S_1$, we see that $y$ is a fast variable
compared to $\delta$. Thus, in this interval of time, we could take $\dot{y}
= 0$, i.e., 
\begin{equation}
\delta = y - \left(\frac{b_0 a z}{\delta}\right). 
\end{equation}
\noindent
Since all these variables $\delta$, $y$ and $z$ are positive on $S_1$, we
see that $y > \delta$. (Note the factor $b_0az/\delta$ is small.) Using this
in Eq. (15) we see that $y$ decreases. Now, consider the equation for $\phi$%
. Using $x \sim {\delta}/{b_0}$ on $S_1$, we get 
\begin{equation}
\dot{\phi} = d \left[ e - \frac{\phi^m \delta}{b_0} \right]. 
\end{equation}

\noindent
Noting the value of $b_0$, we see that $\dot{\phi}$ will be negative when
the trajectory reaches $S_1$. The time scale of evolution of $y$ in Eq. (15)
is of the order of unity while that of $\phi$ is $\sim d/b_0$. These time
scales are relatively fast. (These statements are true only as the
trajectory hits $S_1$.) 
Moreover, since $x$ is a fast variable, the changes in $x$ component
dominates the descent of the trajectory. Finally, as the trajectory
approaches $S_2$, ${\partial g}/{\partial x}$ becomes positive and the
trajectory jumps  from $S_1$ to $S_2$. Combining these results, we see that
the trajectory moves towards the region of smaller values of $y$ and $\phi$
entering $S_2$ in a region of small values of $y$ and $\phi$. \\

\noindent
In summary, the sequential way the orbit visits various parts of the phase
space is as follows. The trajectory enters $S_2$ part of the slow manifold
in regions of small $y$ and $\phi$ making an exit along $S_2$ for relatively
large $\phi$ and $y$. Thereafter, the trajectory moves through the
`unstable' part of the phase space before falling onto the $S_1$ and quickly
descends on $S_1$. This completes the cyclic movement of the trajectory and
explains the geometrical feature of the trajectory shuttling between these
two parts of the manifold and the associated time scales.\\

\noindent
Now, the question that remains to be answered is $-$ do the trajectories
always visit both $S_1$ and $S_2$ or is there a possibility that the
trajectory remains confined to $S_1$ ? It is clear that if the former is
true, relaxation oscillations with large amplitude will occur and if the
latter is true, the oscillations are likely to be of small amplitude. Here,
we recall that the coordinates of  the saddle focus fixed point are $x_0 =
z_0 \sim e/2$ which is much larger than the values of $x$ on $S_2$ ($\sim
y/\left| \delta \right|$). Thus, the fixed point located on the $S_1$ will
be close to the `fold' at the first Hopf bifurcation, $e = e_f$, since the 
latter occurs at small values of $e$ ($e_f \sim 5$). Due to the unstable
nature of the fixed point, the trajectories spiralling out are forced onto
the $S_2$ part of the manifold resulting in relaxation oscillation. This
point has been illustrated by considering the example of a period eleven
orbit for $m=1.2$ and $e=267.0$ shown in Fig. 8. As is clear from this
diagram, the small amplitude oscillations are located on the $S_1$. As the
small amplitude oscillations grow, the relaxation nature does not manifest
until the orbit crosses over to $S_2$. To the best of the authors knowledge,
the mechanism suggested here for pulsed type relaxation oscillations is new. 
\\

\noindent
The above feature of the trajectories continuing in the same direction of
the slow manifold ($S_2$) well into the `unstable' part of the phase space
is somewhat similar to {\it canard solutions} where the trajectories tend to
follow the slow manifold well into the repulsive part of the slow manifold
before jumping to a attracting branch \cite{mil98,eck83}. The differences,
however, are clear. While in canard solutions, the trajectory tends to move
along the repulsive part of the slow manifold before jumping to the
attracting branch of the slow manifold, in our case, the trajectory leaves
the slow manifold and moves into the `unstable' part of the phase space
which is not a part of the slow manifold. \noindent

\subsection{Mixed Mode Oscillations}

\noindent
We now consider the origin of the MMO sequences in our model. Global
bifurcation scenarios are known to be relevant to the MMOs and in the
introduction, we briefly mentioned two of the possible global bifurcation
scenarios which display MMO like sequences. These scenarios are based on the
homoclinic contacts of an equilibrium set like the saddle focus fixed point
and saddle periodic orbit for the Shilnikov and Gavrilov-Shilnikov scenarios
respectively. Each of these scenarios are characterized by the bifurcation
diagrams obtained from the stability analysis of the homoclinic orbits and
by the corresponding scaling relations involved in the approach to
homoclinicity \cite{guc90,gas87}.\\

\noindent
First, we consider the similarities of the behavior of our model with the
characteristic features of the Shilnikov scenario. In three dimension, the
Shilnikov criterion is stated in terms of the two possible combinations of
the dimensions for the invariant manifolds of the saddle-focus; the unstable
manifold is two dimensional and stable manifold is one dimensional and vice
versa. For these two cases, the presence of an homoclinic orbit is given by
the condition $|\rho /\lambda |<1$, where the eigen values of the fixed
point are given by $\rho \pm i\omega ,-\lambda $, where $\rho >0,\lambda >0$
for the first case and $\rho <0,\lambda <0$ for the latter. The analysis of
the Shilnikov scenario shows that in the neighborhood of the homoclinic
point, the parameter space is organized such that the period of the
principal periodic orbit tends to infinity as the parameter approaches the
value corresponding to the homoclinic point. In our case, the system is four
dimensional, with the unstable manifold of the fixed point characterized by
a pair of complex eigenvalues $\rho \pm i\omega $ ($\rho >0$) and the stable
manifold by two eigenvalues $\lambda _1<0$ and $\lambda _2<0$. Here, $%
\lambda _1$ stays close to zero and $\lambda _2$ is substantially negative.
Thus, in our case, the criterion $|\rho /\lambda |<1$ refers to $\left| \rho
/\lambda _1\right| <1$. We find that this condition is satisfied only in a
small region just prior to the disappearance of the PPO in a Hopf
bifurcation. A typical plot of the eigenvalues for $m=1.2$ is shown in Fig.
9, where we have also shown the phase $\omega $. Even though $\left| {\rho }/%
{\lambda }\right| <1$ is not satisfied over large portion of $e$ and $m$, we
do see that the period of the periodic orbits tend to increase as $e$ is
increased ($m<2.0$) which is typical of the Shilnikov scenario. A plot
showing the period (of the superstable orbits) verses the deviation from
estimated homoclinic point($e^{*}$) is displayed in Fig. 10 for $m=1.4$. (We
have plotted points from period three onwards.) Here, we have taken the
value of $e^{*}$ to be the value of $e$ for the onset of the last observed
periodic orbit with period $12$ ($e=247.63$). It must be stated that we do
not face any difficulty in locating any of the periodic orbits upto the
period $12$. However, {\it we are unable to detect the next period which we
interpret as an incomplete approach} to the true homoclinic point. Here, it
must be mentioned that incomplete approach to homoclinicity is quite common%
\cite{pet92,gyo92,arn93,kop95,mil98,kop92a,kop92b,cha95,hau96}. We stress
that even in the region where Shilnikov criterion is obeyed, we do not
observe homoclinic orbit.\\

\noindent
One other feature which is usually seen in the Shilnikov scenario is that
the reinjection of the trajectory in to the neighborhood of the fixed point
is along the direction of the fast variable after which the trajectories
tend to stay around the saddle focus fixed point. In our model, since $x$ is
the fastest variable, it also acts as a reinjection direction. However, a
closer examination shows that the spiraling in of the orbit towards the
fixed point is along the $z$ direction which is the next fastest variable.
This is evident in Fig. 11 where a typical trajectory is shown. Even more
dominant feature of the Shilnikov scenario is that the successive
bifurcations should be connected by the PPO. This, however, is not true in
our case as we have seen earlier, since the isolas which form the period
adding sequence are distinct from the PPO. This feature is clear from Fig. 3
for $m=1.8$. In fact, this feature of the isolas being distinct from the PPO
is more like that of the Gavrilov-Shilnikov scenario which requires the
presence of an unstable periodic orbit {\it which we failed to detect} in
the entire parameter region wherein nontrivial dynamics is present. This
rules out the possible presence of any homoclinic bifurcation due to the
saddle periodic orbit. Thus, we see that our model has partial features of
both these scenarios.\\

\noindent
The above discussion suggests that the origin of the MMOs in our model is
likely to be different from the two scenarios. In order to understand the
mechanism of the MMOs in our model, we will use the information on the
nature of the relaxation oscillations. We first note that the fixed point $%
(x_0,y_0,z_0,\phi _0)$ is on the $S_1$ part of the slow manifold and moves
up on $S_1$ as $e$ is increased. Both $x_0$ and $z_0$ have a near linear
dependence on $e$ namely $x_0=z_0\sim \frac e2$, while $y_0$ and $\phi _0$
are practically constant. Since the fixed point is unstable, any orbit in
its neighborhood will locally expand along its unstable directions. Thus, we
expect to get insight into the mechanism of the MMOs by studying the rate of
expansion of such orbits. In order to understand the mechanism operating in
our model, let us consider a periodic orbit of $L^s:1^{10}$ type shown in
Fig. 12. If the orbit has reached the neighborhood of the fixed point, any
orbit on $S_1$ should spiral out with a local dynamics determined by the
linearized eigenvalues around the fixed point. Within this approximation,
the orbit expands at rate $exp[2\pi \rho /\omega ]$ per rotation around the
fixed point. Assuming a linearized behavior for $n$ rotations ( for fixed
values of $e$ and $m$), we get 
\begin{equation}
r_n/r_1=\exp [2\pi (\rho /\omega )(n-1)]
\end{equation}
where $r_n$ is the distance measured from the fixed point after $n$
rotations along a fixed direction in the unstable manifold of the fixed
point. Here, $r_1$ is the value of $r_n$ for $n=1$. Since Eq. (19) is based
on linearized approximation, the values of $r_n$ obtained from the phase
plots will be in general different due to the influence of nonlinearities.
For this reason, we will first study the region of validity of Eq. (19). We
note that the unstable manifold is nearly in $x-z$ plane and in the
neighborhood of the fixed point, the major contribution to $r_n$ comes from $%
x$ and $z$. Thus, it would be sufficient to consider $x(t)$ (or $z(t)$) in
place of $r_n$ and in particular, we will use the minima or maxima to
analyze the small amplitude oscillations of the periodic orbit using Eq.
(19). Consider the plot shown in Fig. 12. We note that the time interval
between the successive minima or maxima of $x(t)$ can be taken to correspond
to one rotation of the orbit. We shall denote the the deviations of the $n$%
-th minimum from the fixed point value, $x_0(e)-x_n^{min}$, by $x_n^{*}$. In
Fig.12, $x(t)$ and $z(t)$ are plotted along with their fixed point values ($%
x_0=z_0=137.82$). ( We have also shown regions of $x(t)$ corresponding to
regions $S_1$ and $S_2$ of the slow manifold.) The decreasing nature of the
maxima values of the amplitude of $x(t)$ for the first few cycles is a
reflection of the fact that the orbit has not reached the neighborhood of
the unstable manifold of fixed point lying on $S_1$ (see Fig. 11). It is
only later (in time) that the expanding nature of the oscillations manifest,
seen as the increase in the magnitude of the successive maxima values of $%
x(t)$. Thus, the value of $x_1^{*}$ read off from $x(t)$ (in place of $r_1$)
will contain contributions arising from reinjection mechanism. Hence,
identifying $x_1^{*}$ as representing the value of the first minimum of $x(t)
$ would be an incorrect, if one wishes to use Eq. (19). Thus, $x_1^{*}$ has
to be estimated by extrapolating the values of $x_n^{*}$ using values of $n$
where nonlinearity plays an insignificant role, {\it i.e.}, $n>3$ for the
case of Fig.12. We denote this extrapolated value by $x_1^{\dagger }$. In
addition, as the amplitude grows as a function of $n$, one should also
expect that the linear dynamics breaks down. Thus, for larger $n$ values, we
should again see the effect of nonlinearity. In fact, this feature shows up
as a decrease (though marginal) in the time interval between successive
minima for higher $n$ as can be verified from Fig. 12 (seen between the
ninth and the tenth peaks).\\

\noindent
Now, we attempt to estimate the {\it changes} in the magnitude of the small
amplitude oscillations located on $S_1$ as a function of $e$ and estimate at
what value of $e$ the trajectory hits the `fold' between $S_1$ and $S_2$. We
denote the distance of the fixed point $(x_0 (e), y_0 (e), z_0 (e), \phi_0
(e))$ from the `fold' given by $D=((x_0(e)-x_0(e_f ))^2 +
(y_0(e)-y_0(e_f))^2 + (z_0(e)-z_0(e_f))^2 + (\phi_0(e)-\phi_0(e_f))^2)^{%
\frac{1}{2}}$, where $(x_0(e_f),y_0(e_f),z_0(e_f),\phi_0(e_f))$ refers to
the value of the fixed point at the first Hopf bifurcation ($e=e_f$).
Further, we note that the fixed point is close to the `fold' at the first
Hopf bifurcation and noting $x_0=z_0\sim\frac{e}{2}$ with a very weak
dependence of $y_0$ and $\phi_0$ on $e$. Thus, in one dimension where we are
dealing with $x$ variable alone, we can take $D \sim x_0(e)-x_0(e_f)$. (Note
that position of `fold' is insensitive to $e$.) Using the fact that $x_0(e_f)
$ is small, we get $D \sim x_0(e)$. Using this, we will attempt to find the
maximum value of $n$ allowed for which the condition $x_n^* > D$  is
satisfied. For this, we need to know the dependence of $\rho$ and $\omega$
on $e$ which decides the rate of growth of the small amplitude oscillations
as a function of $e$. In the range of the MMO sequences that we are
interested, $\rho$ decreases and $\omega$ increases nearly linearly as can
be seen from Fig. 9. Thus, we take $\rho = \rho_0 - m_{\rho} e$ and $\omega
= \omega_0 + m_{\omega} e$, where, $m_\rho$ and $m_{\omega}$ are the
corresponding slopes. These are evaluated numerically as the best fit for
the region of interest of $e$. The fit yields $\rho_0=0.10433, m_\rho =
0.0003632, \omega_0\approx 0$ and $m_\omega=0.0004114$ for the case shown in
Fig. 12. Using this in Eq. (19) gives 
\begin{equation}
\frac{x_n^*}{x_1^*} = \exp [ 2\pi (\rho/\omega) (n-1)] = \exp \{ [k_1 +
k_2/e] (n-1) \} 
\end{equation}
\noindent
Here $k_1$ and $k_2$ are functions of $\rho_0$, $m_\omega$ and $m_\rho$.
This equation can be interpreted as a scaling form for the small amplitude
oscillations of stable periodic orbits (i.e., for a fixed $n$) as a function
of $e$. Now, consider the set of all stable periodic orbits of the form $L^s$
for a given value of $m$. Then, for each of these $L^s$ orbits, the values
of $n$ ranges upto $s$. Since the magnitude of small amplitude oscillations
of these periodic orbits depends $e$, we can plot $ln\,x_n^*$ versus $1/e$,
where we have used $x_n^*$ values corresponding to the $n$-th minimum of a
periodic orbit. The plots of $ln\,x_n^*$ as a function of $e^{-1}$ for $n=4$
to $7$ are shown in Fig. 13. In the figure, the lowest curve (+) corresponds
to $n=2$ and other successive higher curves refer to $n = 3$ upwards. The
lowest band within the $n =2$ curve corresponds to a periodic orbit of the
form $L^{10}$ and successive bands have decreasing $s$ values. (The gaps
correspond to chaotic bands between successive periodic orbits.) It is clear
that the plots are linear and the slopes of the curves corresponding to $n =
4,5,6$ and 7 show an increasing trend, increasing in multiples of $1530$ ($%
k_2$). To illustrate the presence of nonlinearity, we have also shown plots
of $ln\,x^*_n$ for $n = 2$ and 3. One can easily notice that these two
curves deviate from linearity considerably for large $e$. In addition, we
see that the slopes of these two lines are not in multiples of $k_2$. This
suggests that it should be possible to collapse all the curves for $n > 3$
onto a single curve. Noting that the slopes of curves are in multiples of $%
k_2$, $ln x_n^*/n$ renders them parallel and the preexponential factor which
satisfies Eq.(20), denoted by $x_1^\dagger$ can be easily determined. The
value of $x_1^\dagger$ so obtained can be now used in 
\begin{equation}
\frac {x_n^*} {x_1^\dagger} = \exp [ 2\pi (\rho/\omega) (n-1)] 
\end{equation}
to estimate the maximum number of small amplitude oscillations allowed for
any given value of $e$ before the size of the orbit (under the linear
approximation) is large enough to hit the `fold'. This is determined by the
value of $n = n_c$  at which $x_n^* = x_{n_c}^* > D \sim e/2$.\\

\noindent
In the inset of Fig. 13, we have shown that the curves for $n=4$ to $7$ can
be collapsed on to a single curve. The value of $x_1^{\dagger}$ obtained by
extrapolating the curves $ln\,(x_n^* (e))/n$, for $n=7$ to $4$ is  5.617. We
have verified that the $n=2$ curve and to a lesser extent $n=3$ curve
deviates from the collapsed curve reflecting that nonlinearity corresponding
to reinjection is dominant for these two cases. This also implies that the
orbits do not approach the fixed point close enough that the linearized
eigenvalues could be useful.  Using the value of $\rho = 0.0073$ and $\omega
= 0.1348$ for $e=267.0$, we can now estimate the value of small period
oscillations for $n =11$ is 171.0 which is larger than $e/2$. (Here $e =
267.0$.) This means that a maximum of {\it nine small amplitude oscillations}
are allowed {\it beyond the first minimum} at this value of $e=267.0$. This
is consistent with what is seen in Fig. 12. We have verified that this
method of estimating the maximum number of allowed small amplitude
oscillations for any given value of $e$ works very well as long as $n > 3$
for $m = 1.2$ and other values of $m$ as well. For $n=2$ and 3, the value of 
$e$ at which these orbits disappear shifts to much lower values that what
actually observed.\\

\noindent
A little reflection on the above results reveal the cause of incomplete
approach to the homoclinic point. Recall that the growth of small amplitude
oscillations is controlled by $2\pi \rho /\omega $. We have seen that this
quantity depends inversely on $e$. Thus, the arithmetically increasing
number of small amplitude oscillations accommodated on $S_1$ ( without the
trajectory crossing over to $S_2$) is a direct result of `softening' of $%
2\pi \rho /\omega $ as a function of $e$. In other words, for every small
amplitude oscillation accommodated on $S_1$, $e$ changes by a fixed amount
commensurate with the softening rate. Thus, the number of small amplitude
oscillations that can be accommodated in the allowed interval of $e$ in the
bifurcation plot will be limited. This also implies that the approach to
homoclinic point can at best be asymptotic due to finite rate of softening
of $2\pi \rho /\omega $, with the asymptotic nature manifesting only in the
limit $k_2\rightarrow 0$. Clearly, these results are valid under the
assumption that the contribution from nonlinear terms to the growth of the
small amplitude oscillations is not strong, which is substantiated by the
numerical evidence that the estimated number of small amplitude oscillation
allowed for a particular value of $e$ agrees with what is numerically
observed.\\

\noindent
It may be worth pointing out here that even though the analysis given here
is for MMOs of the kind $1^s$, it is clear that they can be easily
generalized to $L^s$ kind of MMOs. It must be mentioned here that the
`softening' of $2\pi \rho/\omega$ is a result of the global constraint in
our model, namely, the back-to-back Hopf bifurcation. Thus, we see that the
apparent homoclinic scenario exhibited by the model system is completely
new. We have shown that this feature coupled with the mechanism of the
relaxation oscillations operating in the model gives rise to features of
MMOs common to both Shilnikov and Gavrilov-Shilnikov scenario.\\

\subsection{Discussion and Conclusions}

\noindent
Some comments may be in order here about the bent-slow manifold structure of
the system. The relaxation oscillations seen in this system differs
qualitatively from that seen in systems with the $S-$shaped slow manifold.
The major difference in the structure is that while in the $S-$ shaped
manifold, there are two attractive pleats separated by a repulsive part, in
our case, both pieces $S_1$ and $S_2$ of the bent-slow manifold are
attractive and are connected continuously. This aspect coupled with the fact
that there is no repulsive part in the bent-slow manifold as in the $S$-
shaped manifold suggests that the mechanism causing jumps between the $S_1$
and $S_2$ is very different. The only similarity is that the number of jumps
( fast transitions) accomplished by the trajectories from one part of the
slow manifold ($S_2$) to another ($S_1$) and vice versa is two as in the
case of $S-$shaped manifold. During the jump from $S_2$ to $S_1$, the
trajectory tends to move out of the slow manifold into unstable phase space
by sticking to the direction of of motion on $S_2$. Here the motion is
accelerated due to unstable nature of the phase space. This has some
similarity to {\it canard } type of solutions but the comparison is
superficial since the unstable part of the phase space to which the
trajectory moves is not a part of the slow manifold.  There is another
difference namely the operative time scales in the dynamics on the slow
manifold. The dynamics on $S_2$ is slow as it is controlled by the slow
variables $y$ and $\phi$, since $\dot x \sim 0$ for the entire interval of
time the trajectory is on $S_2$. On the other hand, on $S_1$, the time
dependence of a trajectory is largely controlled by the fast variable $x$. \\

\noindent
We have also analyzed the effect of bent-manifold structure on mixed mode
oscillations and the incomplete homoclinic scenario. We have shown that in
the case of bent-manifold structure, the approach of the fast variable
towards the fixed point is along the slow manifold itself eventhough the
eventual approach is along the $z$ direction. This is in direct contrast
with the $S-$shaped structure where the direction of the jump of the fast
variable is transverse to the slow manifold containing the fixed point as in
the Rossler's $S-$shaped slow manifold. There is another possibility as
pointed out by Koper {\it et al}\cite{kop92b}. In this case, the plane of
relaxation oscillations is parallel to the plane of nearly harmonic small
amplitude oscillations. These authors suggest that this type of dynamics may
be responsible for the incomplete approach to homoclinicity since the
approach to the fixed point is along the attracting pleat of the $S-$shaped
manifold. However, in both cases, irrespective of the location of the fixed
point, the nature of the relaxation oscillations remain the same. Moreover,
in both these cases, the small amplitude oscillations(near harmonic), as
well as the large amplitude (relaxation type) oscillations are well
characterized. However, the nature of intermediate amplitude oscillations
are not so well understood. It has been suggested in the literature that
these are related to {\it canard} type of solutions\cite{mil98,eck83}. The
latter type of oscillations result from `sticking' of the trajectory to the
repelling part of the $S-$shaped slow manifold before jumping to the
attracting pleat of the slow manifold. In our case, although the
oscillations have a superficial similarity with {\it canard} type of
solutions, it is the 'sticking' of the trajectories in the same direction of 
$S_2$ well into the unstable phase space, coupled with the fact that there
is no inherent constraint relating $S_1$ to $S_2$ in the manifold structure
that appears to lead to jumps of all sizes. As an illustration, we have
shown a plot of the trajectory for $m=1.8$ and $e=190.0$ in the $x-\delta $
plane (Fig.14). It is clear that while the small amplitude oscillations are
located in the neighborhood of the fixed point, the intermediate amplitude
oscillations result from the trajectory `sticking' to the direction of the $%
S_2$ plane and moving into the `unstable' part of the phase space by varying
amounts each time the trajectory leaves $S_2$.\\

\noindent
Although the bent-manifold structure is characteristic of our systems, we
believe similar structure is likely to be seen in many other models and
experimental systems. Particularly, in chemical kinetics where only binary
collisions are permitted, models are expected to involve only quadratic or
biquadratic nonlinearities. Such models are promising candidates to exhibit
the bent-slow manifold structure. In experimental systems, the dominant
signature to look for would be the trapping of fast variable at small values
over a substantial portion of its period followed by sharply peaked pulse
like behavior.\\

\noindent
From the above discussion, we see that the bent-slow manifold structure is
at the root of understanding of the pulsed type of relaxation oscillation
and the MMOs. We note here that our analysis is completely local since we
have used the linearized eigenvalues around the fixed point. For the same
reason, the scaling relation obeyed by the small amplitude oscillations (Eq.
21 and Fig. 13) is found to be valid only where the influence of
nonlinearity to the growth of these small amplitude oscillations (as a
function of $e$) is minimal. In spite of this, the scaling relation so
obtained forms the basis for estimating the maximum number of small period
oscillations permitted for a given $e$ thus explaining the origin of MMOs in
the model. \\

\noindent
Here, we mention that the relaxation oscillations arising out of the
atypical bent-slow manifold structure is directly related to a dominant
characteristic of the PLC effect, namely, the negative strain rate
sensitivity of the flow stress. The analysis of time scales involved in the
relaxation oscillations has been useful in understanding the origin of the
negative strain rate sensitivity of the flow stress which is reported
elsewhere \cite{raj99}.\\

\noindent 
In summary, we have analyzed the dynamics of a model for a type of plastic
instability due to Ananthakrishna and coworkers with particular attention to
the complex dynamics exhibited by the model. We have shown that the nature
of relaxation oscillations and the MMO sequences exhibited by the model is
atypical. We have proposed a new mechanism for the relaxation oscillations
based on the bent-slow manifold structure of the model. Using this we have
explained the origin of the MMOs. We have further shown that a crucial role
in organizing the dynamics is played by the physical constraint, namely, the
stress oscillations are seen only in a window of strain rates. (Indeed, the
model has been devised to be consistent with this experimental feature.)
This constraint translates to back-to-back Hopf bifurcation in the model
leading to the `softening' of the eigenvalue of the saddle fixed point that
controls the small amplitude oscillation. It is this finite rate of
softening that is responsible for the incomplete approach to the homoclinic
bifurcation.

\section{Acknowledgements}

\noindent
The authors would like to thank Dr.T.M. John who was involved in the early
stages of the work and one of the authors (SR) wishes to thank the theory
group, Materials Research Center, IISc, for helpful discussions.

\newpage
\begin{center}

\begin{figure}
\centering
\mbox{
\epsfxsize=14cm \epsfbox{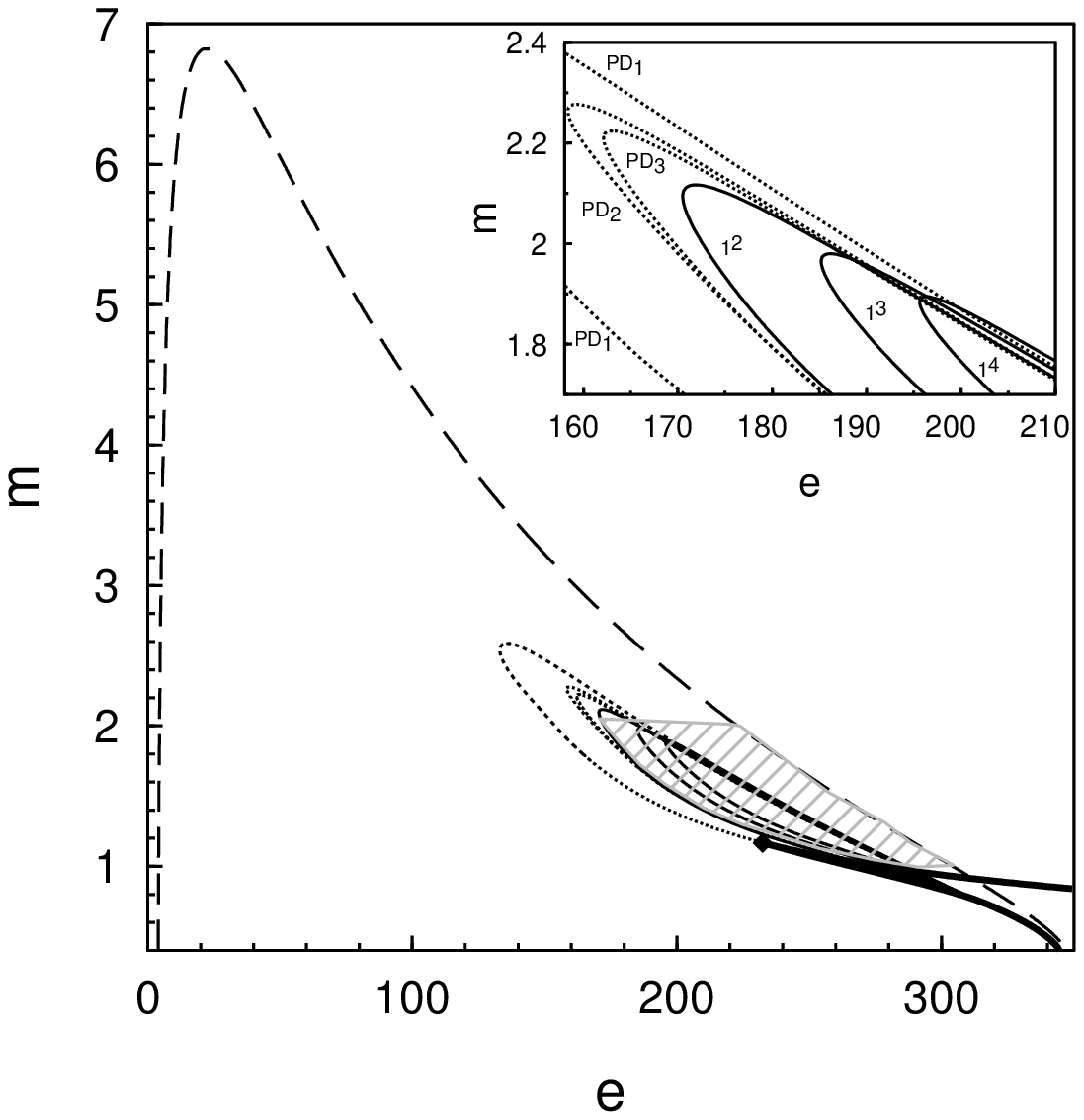}}
\caption{Phase diagram of the model in $(m,e)$ plane.
The broken line corresponds to the locus of Hopf bifurcations, dotted
lines  to the PD bifurcations and the continuous lines to the
locus of SN bifurcations.  The thick lines represent  the  SN bifurcations
of the PPO culminating
in a codimension 2 cusp bifurcation point shown as filled diamond.
Approximate region of MMOs is  shown by the hatched area.}
\end{figure}

\begin{figure}
\centering
\mbox{
\epsfxsize=12cm \epsfbox{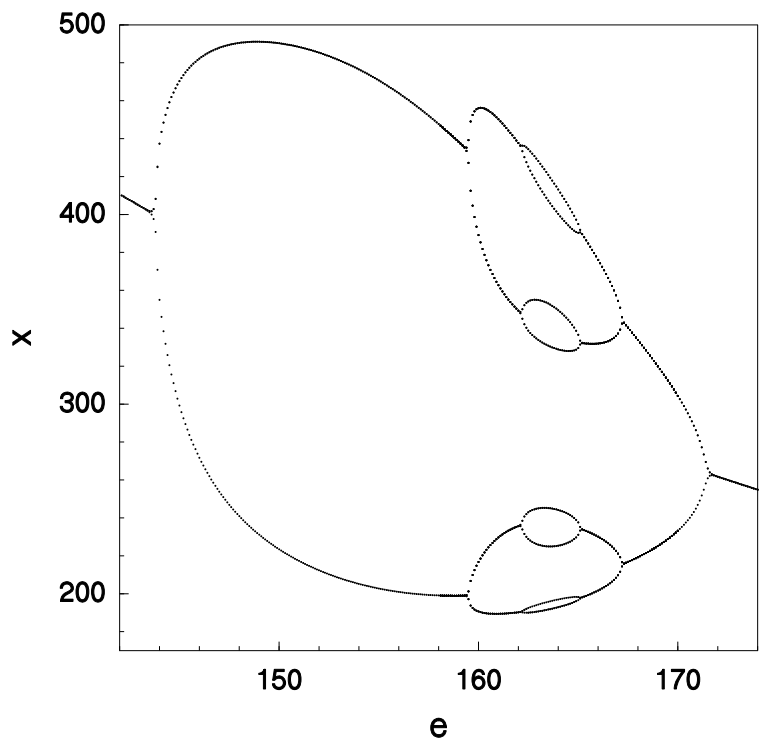}}
\vspace{-1cm}
\caption{
Bifurcation diagram for $m=2.215$.}
\end{figure}

\begin{figure}
\vspace{-2cm}
\centering
\mbox{
\epsfxsize=12cm \epsfbox{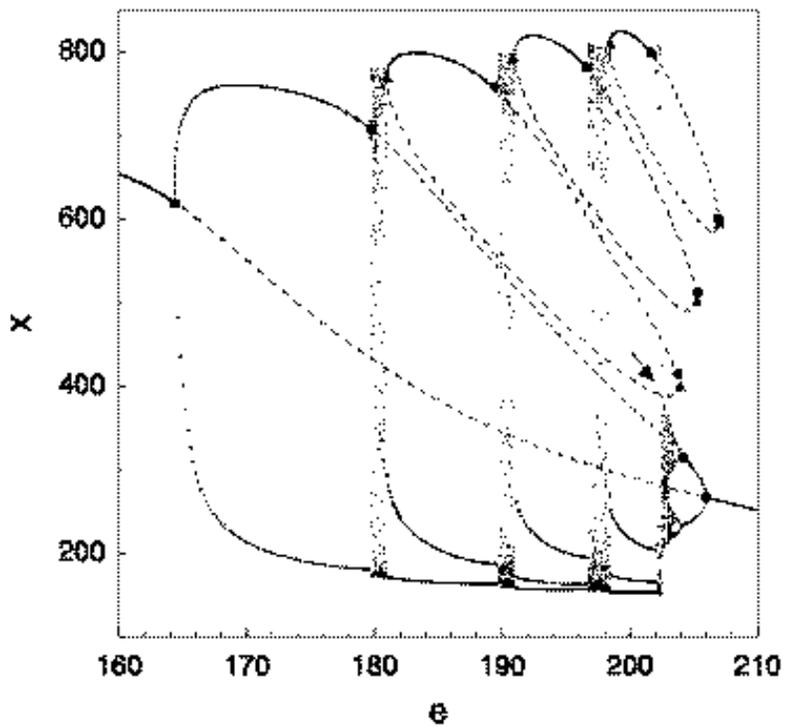}}
\vspace{-1cm}
\caption{
The bifurcation diagram for $m=1.8$.  Dashed
lines indicate  unstable periodic orbits.  Filled circle correspond to
a PD bifurcation and  filled triangle correspond to SN bifurcation. An interior  
crisis point is marked by an arrow.}
\end{figure}

\begin{figure}
\centering
\mbox{
\epsfxsize=12cm \epsfbox{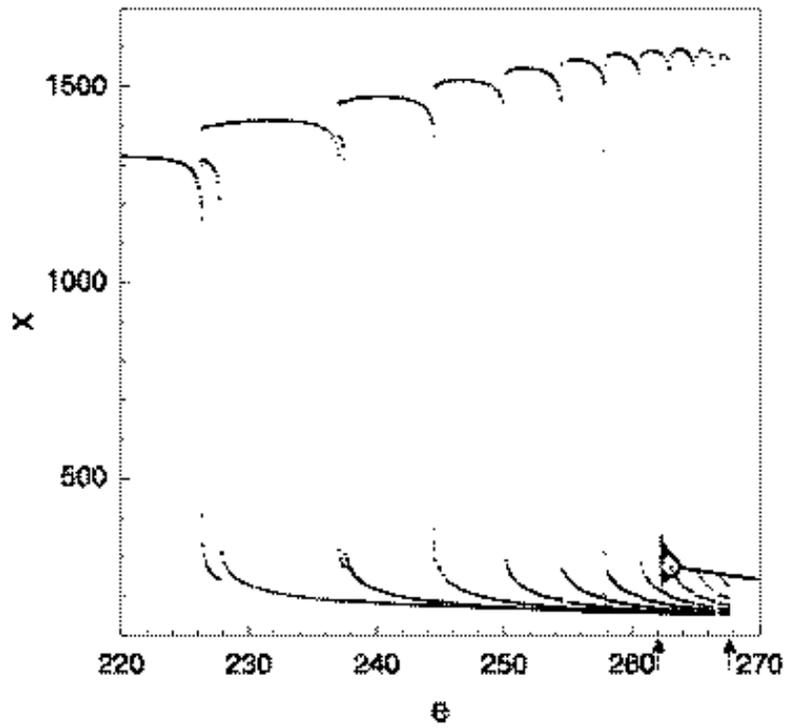}}
\vspace{-1cm}
\caption{
Bifurcation diagrams for $m=1.2$.
The region between the arrows correspond to the   coexistence region of the MMOs
and  small amplitude periodic orbits.}
\end{figure}

\begin{figure}
\centering
\mbox{
\epsfxsize=12cm \epsfbox{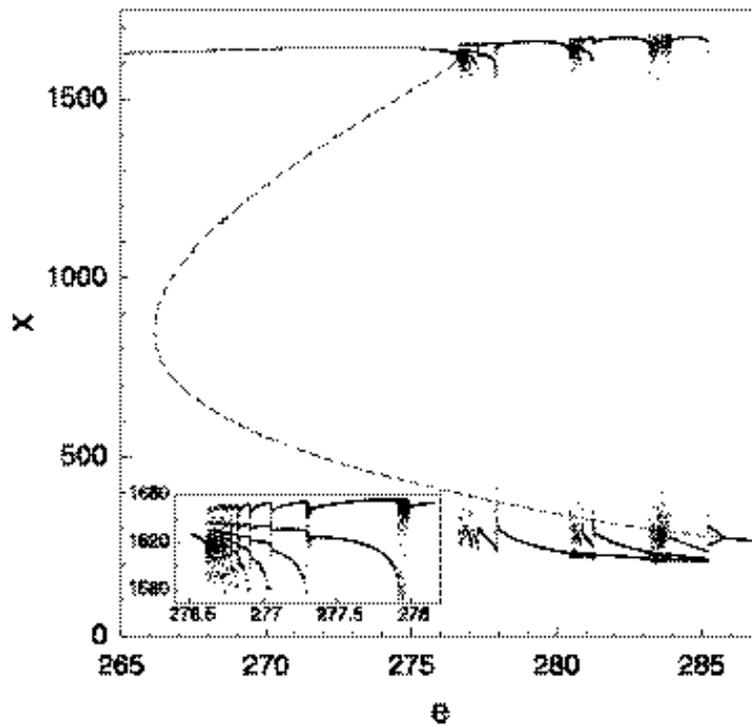}}
\vspace{-1cm}
\caption{Bifurcation diagram  for $m=1.0$.
The dashed line indicates the unstable PPO.
Inset shows enlarged portion of the bifurcation diagram where the 
secondary Farey sequence manifests immediately after the SN bifurcation of the PPO. }
\end{figure}

\begin{figure}
\centering
\mbox{
\epsfxsize=12cm \epsfbox{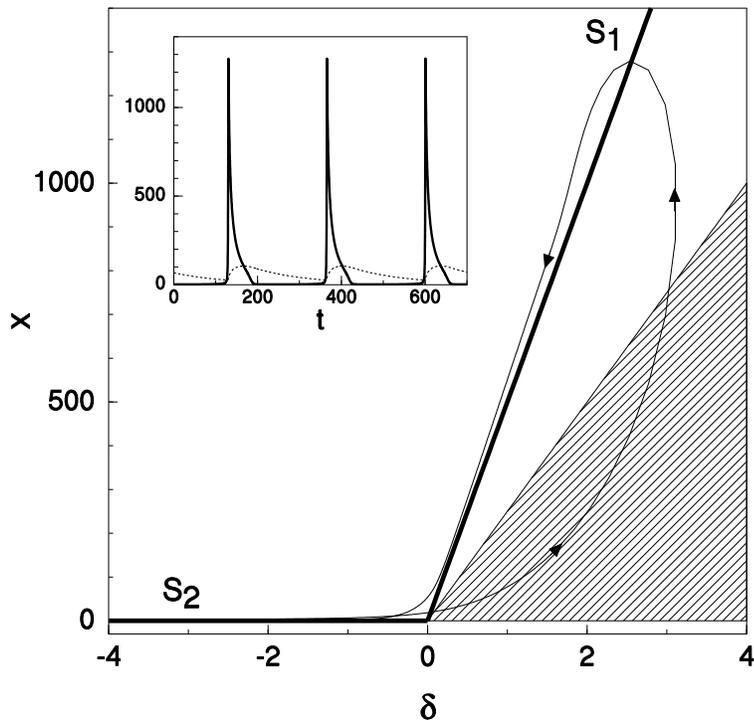}}
\vspace{-1cm}
\caption{
Evolution of a trajectory (thin lines)  along with the  bent-slow
manifold ($S_1$ and $S_2$ shown by thick lines)
structure in the $x-\delta$ plane, for $m=1.2$ and $e=200$.
Inset shows the time series of the $x$ variable (continuous line) and $z$
variable (dotted line).}
\end{figure}

\begin{figure}
\centering
\mbox{
\epsfxsize=12cm \epsfbox{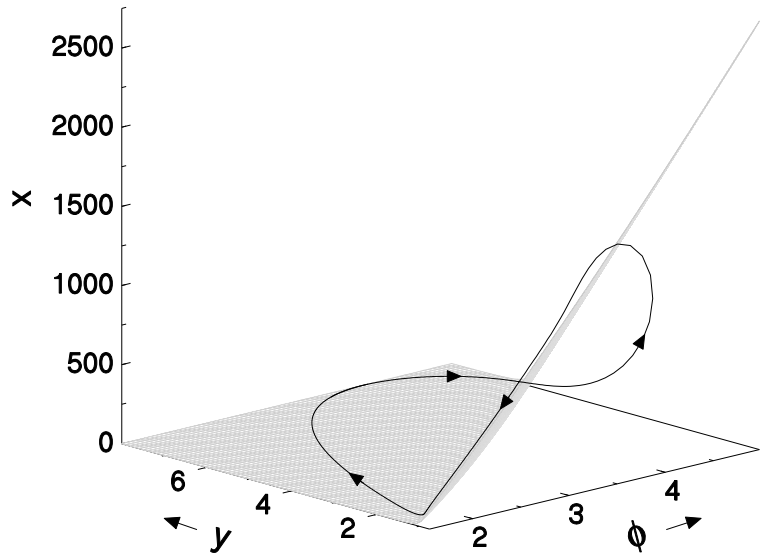}}
\vspace{-1cm}
\caption{
Evolution  of the trajectory along with  bent-slow manifold ($S_1$ and $S_2$)
structure  in ($x,y,\phi$) space indicated by  the gray plane, for $m=1.2$ and
$e=200.0$.}
\end{figure}

\begin{figure}
\centering
\mbox{
\epsfxsize=12cm \epsfbox{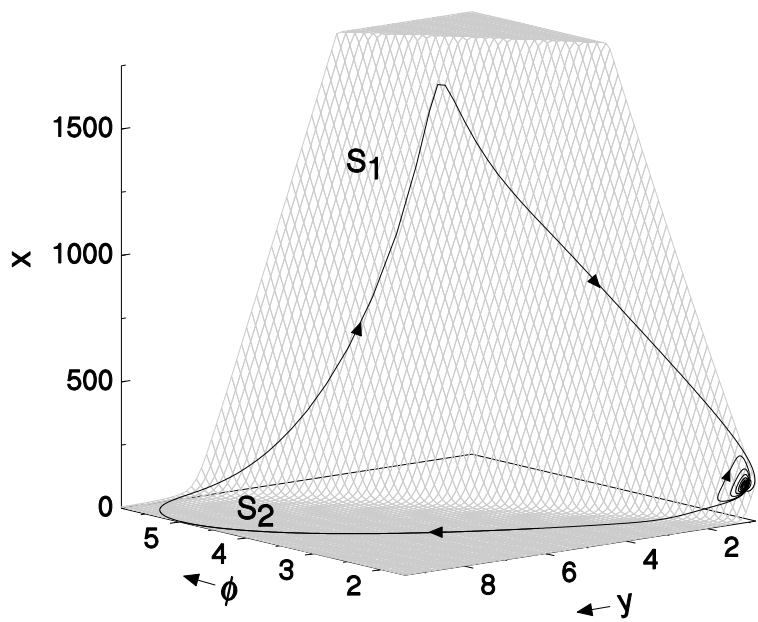}}
\vspace{-1cm}
\caption{
Evolution of the trajectory along  the bent-slow manifold ($S_1$ and $S_2$) structure
for $m=1.2$ and $e = 267.0$.}
\end{figure}

\noindent
\begin{figure}
\centering
\mbox{
\epsfxsize=12cm
\epsfbox{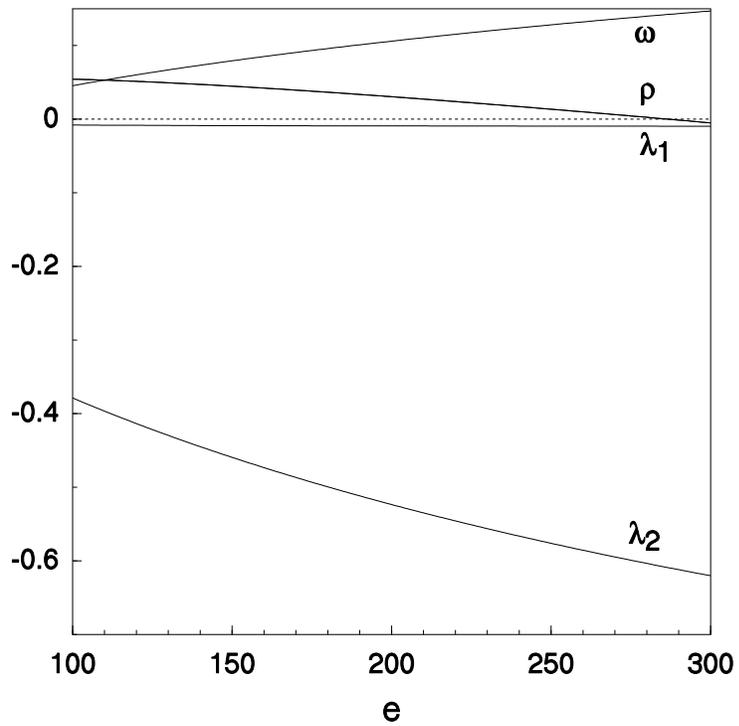}}
\vspace{-1cm}
\caption{
Eigen value spectrum of  the fixed  point for $m=1.2$. The dotted line represents
the zero value.}
\end{figure}

\begin{figure}
\centering
\mbox{
\epsfxsize=12cm
\epsfbox{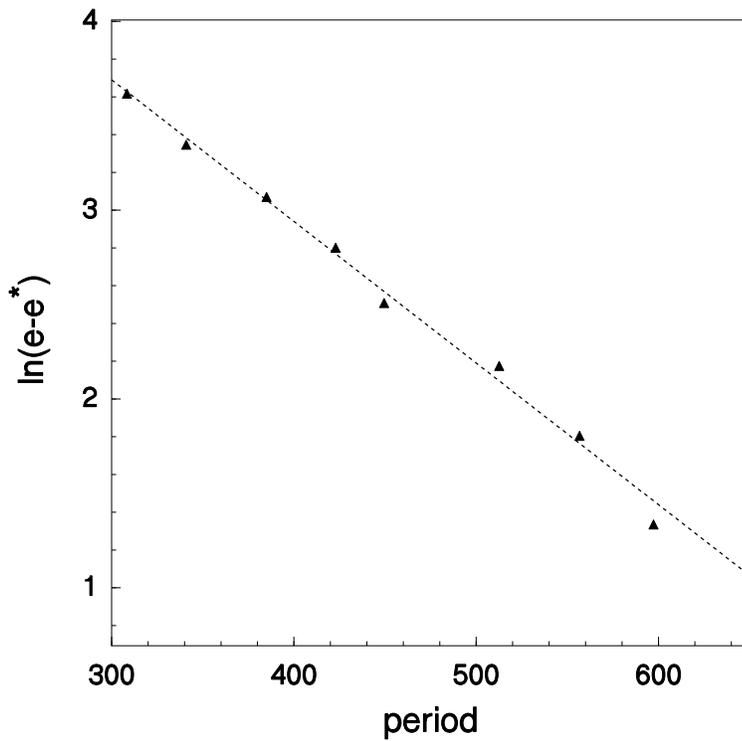}}
\vspace{-1cm}
\caption{
Scaling of   the period of the superstable orbits with the parameter for $m=1.4$.
The  exponential scaling indicates  the apparent  approach to homoclinicity of the
The  exponential scaling indicates  the apparent  approach to homoclinicity of the
saddle focus fixed point. Here $e^*$ is taken to be the value of the twelve
period  orbit, $e = 247.63$.}
\end{figure}

\begin{figure}
\centering
\mbox{
\epsfxsize=12cm
\epsfbox{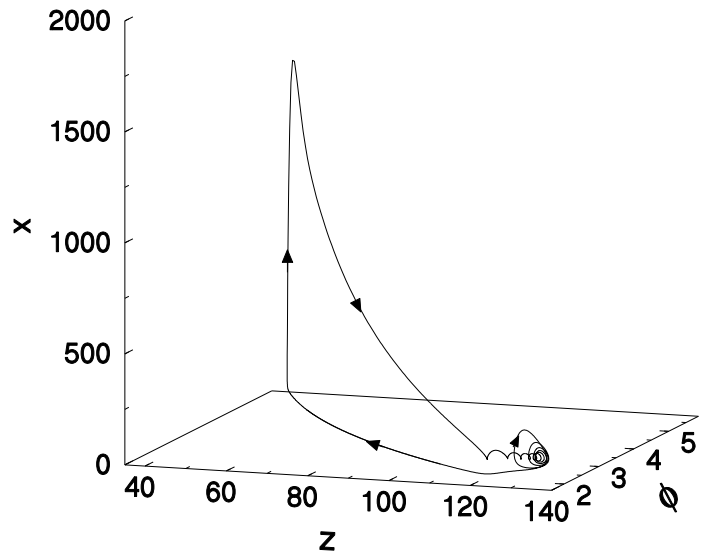}}
\vspace{-1cm}
\caption{
Plot of a periodic orbit in ($x,z,\phi$) space for $m=1.2$ and
$e=267.0$ showing the eventual  
direction of approach  towards the  fixed point is  $z$
direction.}
\end{figure}

\noindent
\begin{figure}
\centering
\mbox{
\epsfxsize=12cm
\epsfbox{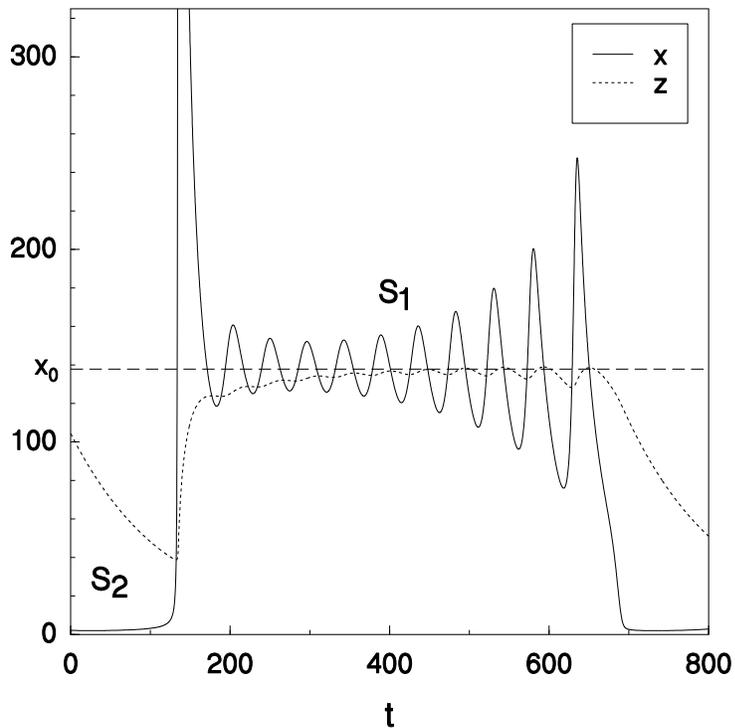}}
\vspace{-1cm}
\caption{Time series of $x$ and $z$ plotted  for $m=1.2$ and
$e=267.0$.  Dashed line shows the fixed point value.}
\end{figure}

\begin{figure}
\centering
\mbox{
\epsfxsize=12cm
\epsfbox{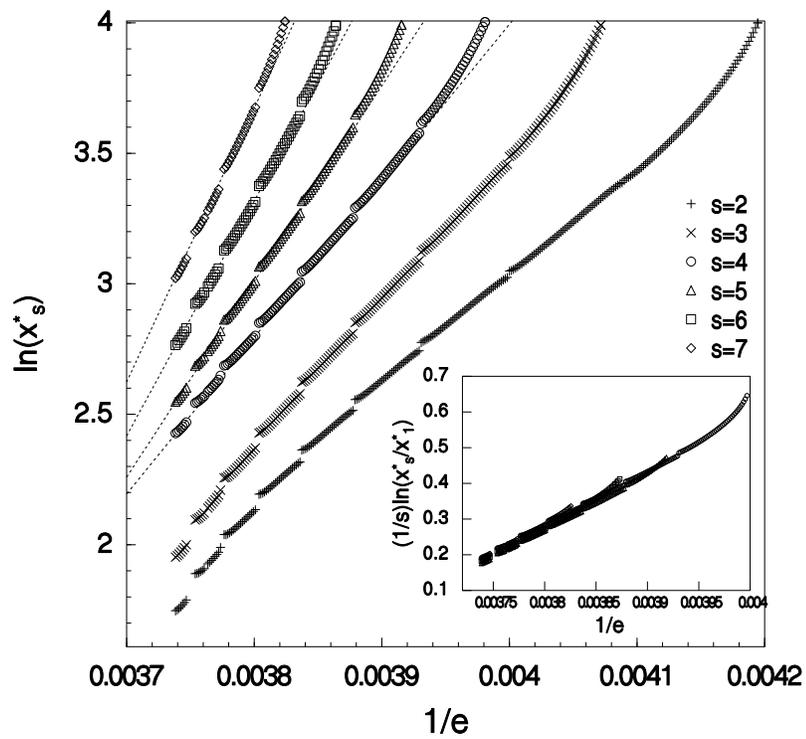}}
\vspace{-1cm}
\caption{
 Scaling of minimum value of $x$ with
$e^{-1}$ for  $n = 2,3, \cdots,7$ ($m=1.2$).
Inset shows  collapse of all the curves from $n=4$ to $7$ onto a single
curve. Dashed lines are shown as guide to the eye.}
\end{figure}

\begin{figure}
\centering
\mbox{
\epsfxsize=14cm
\epsfbox{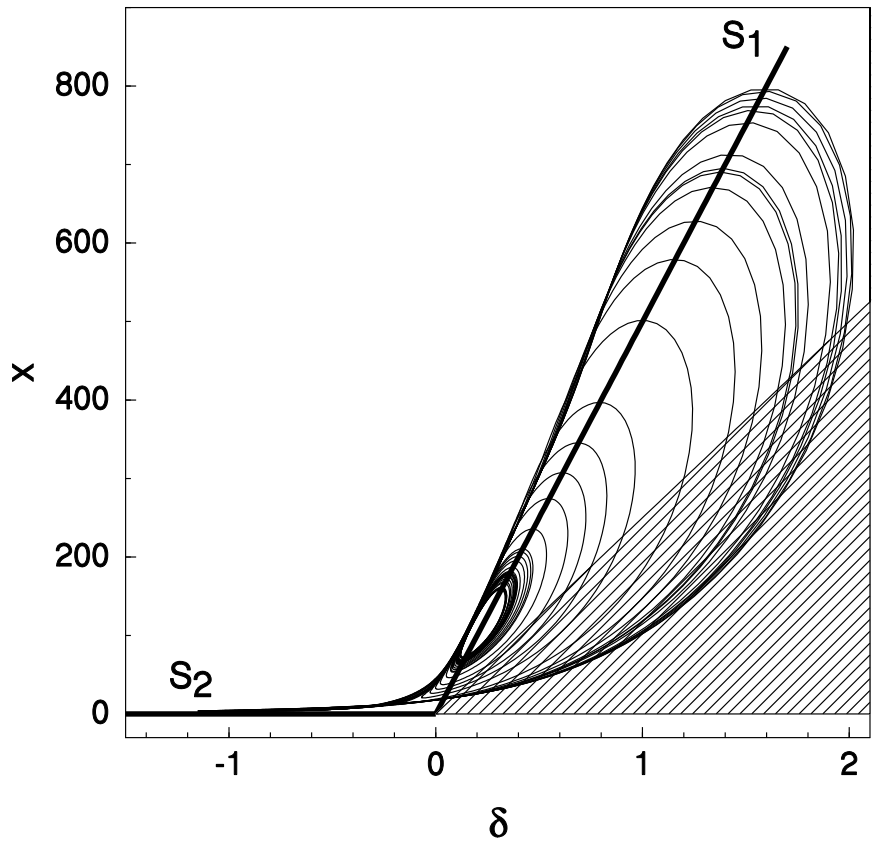}}
\caption{
A chaotic  trajectory in $x-\delta$ plane for $m=1.8$ and $e=190.0$.}
\end{figure}
\end{center}

\end{document}